\documentclass[fleqn,twoside]{article}
\usepackage{espcrc2}
\usepackage{graphicx}
\newcommand{\lesssim}{\mathbin{\lower 3pt\hbox
   {$\rlap{\raise 5pt\hbox{$\char'074$}}\mathchar"7218$}}} 
\newcommand{\gtrsim}{\mathbin{\lower 3pt\hbox
   {$\rlap{\raise 5pt\hbox{$\char'076$}}\mathchar"7218$}}} 
\newcommand{\be}{\begin{eqnarray}}
\newcommand{\ee}{\end{eqnarray}}

\newcommand{\SCN}{S_{\rm CN}}
\newcommand{\DCN}{D_{\rm CN}}
\newcommand{\Ssyso}{S_{\rm sys_0}}

\newcommand{\na}{n_{\rm a}}
\newcommand{\half}{\frac{1}{2}}
\newcommand{\Npol}{N_{\rm pol}}
\newcommand{\Nnu}{N_{\nu}}
\newcommand{\bc}{b_{\rm c}}
\newcommand{\bckm}{b_{\rm c,km}}
\newcommand{\Npix}{N_{\rm pix}}

\title{The Dynamic Radio Sky}
\author{James M.~Cordes\address[Cornell]{Astronomy Department and
		NAIC, Cornell University, Ithaca, NY, USA;
		cordes@astro.cornell.edu}
		\thanks{This work was supported by NSF grants to
		Cornell University, AST~9819931, AST~0138263, and
		AST~0206036 and also by the National Astronomy and
		Ionosphere Center, which operates the Arecibo
		Observatory under a cooperative agreement with the
		\hbox{NSF}.}
	T.~Joseph~W.~Lazio\address[NRL]{Naval Research Laboratory,
		Washington, DC, USA; Joseph.Lazio@nrl.navy.mil}
		\thanks{Basic research in radio astronomy at the NRL
		is supported by the Office of Naval Research.}
	M.~A.~McLaughlin\address[JBO]{Jodrell Bank Observatory, University of Manchester, Macclesfield, Cheshire, SK11 9DL, UK}}

\begin{document}

\begin{abstract}
Transient radio sources are necessarily compact and usually are the
locations of explosive or dynamic events, therefore offering unique
opportunities for probing fundamental physics and astrophysics.
In addition, short-duration transients are powerful probes of
intervening media owing to dispersion, scattering and Faraday rotation
that modify the signals.
While radio astronomy has an impressive record obtaining high time
resolution, usually it is achieved in quite narrow fields of view.
Consequently, the dynamic radio sky is poorly sampled, in contrast to
the situation in the X-ray and $\gamma$-ray bands.
The SKA has the potential to change this situation, opening up new
parameter space in the search for radio transients.
We summarize the wide variety of known and hypothesized radio
transients and demonstrate that the SKA offers  considerable power in
exploring this parameter space.
Requirements on the SKA to search the parameter space include the
abilities to
(1)~Make targeted searches using beamforming capability;
(2)~Conduct blind, all-sky surveys with dense sampling
of the frequency-time plane in wide fields;
(3)~Sample the sky with multiple fields of view from spatially
well-separated sites in order to discriminate celestial and terrestrial
signals; 
(4)~Utilize as much of the SKA's aggregate collecting area
as possible in blind surveys, thus requiring a centrally condensed
configuration;
and 
(5)~Localize repeating transient sources
to high angular precision, requiring a configuration with long baselines,
thus requiring collecting area in both a centrally condensed ``core''
array and sufficient area on long baselines.
\end{abstract}

\maketitle

\section{Introduction}\label{sec:drs.intro}

Transient emission---bursts, flares, and pulses on time scales of
order 1~month---marks compact
sources or the locations of explosive or dynamic events.  As such,
radio transient sources offer insight into a variety of fundamental
physical and astrophysical questions including
\begin{itemize}
\item The mechanisms of efficient particle acceleration;
\item Possible physics beyond the Standard Model;
\item The nature of strong field gravity;
\item The nuclear equation of state;
\item The cosmological star formation history;
\item Detecting and probing the intervening medium(a); and
\item The possibility of extraterrestrial civilizations.
\end{itemize}

A figure of merit for transient detection is $A\Omega(T/\Delta t)$,
where $A$ is the collecting area of the telescope, $\Omega$ is the
solid angle coverage in the search for transients, $T$ is the total
duration of observation, and~$\Delta t$ is the time resolution.
Effective detection of transients requires
\begin{equation}
A\Omega\left(\frac{T}{\Delta t}\right) \to \mathrm{``large\char'42}
\label{eqn:drs.fom}
\end{equation}
At high energies (X- and $\gamma$-rays), detectors with large solid
angle coverage and high time resolution have had great success in
finding classes of transient objects.  At optical wavelengths there
has been recent progress in constructing wide-field detectors with
high time resolution.  Historically, radio telescopes have been able
to obtain high time resolution (with some modern telescopes achieving
nanosecond time resolution) or large fields of view, but rarely have
both high time resolution and large field of view been obtained
simultaneously.

Nonetheless, there are a number of indications that the radio sky may
be quite dynamic.  Radio observations of sources triggered by
high-energy observations (e.g., radio observations of gamma-ray burst
afterglows), monitoring programs of known high-energy transients
(e.g., radio monitoring of X-ray binaries), giant pulses from the Crab
pulsar, a small number of dedicated radio transient surveys, and the
serendipitous discovery of transient radio sources (e.g., near the
Galactic center) suggest that the radio sky is likely to be quite
active on short time scales.  There also may be unknown classes of
sources.

In this chapter, we discuss radio transients from the standpoint of
the available phase space that the SKA can probe and how the design of
the SKA can be optimized to improve its study of both known and
suspected classes of transient radio sources.  Several specific
classes of radio transients---including pulsars, supernovae and
gamma-ray bursts, ultra-high energy cosmic rays, and
scintillation-induced variability---are discussed in more depth in
additional chapters.  We summarize the wide variety of both known and
hypothesized radio transients in \S\ref{sec:drs.sources}, the
available phase space that the SKA can probe in
\S\ref{sec:drs.phasespace}, and the operational modes and aspects of
transient detection for the SKA in \S\ref{sec:drs.survey}.

\section{Known and Potential Classes of Transient Radio
	Sources}\label{sec:drs.sources}

Searches for radio transients have a long history, and a wide variety
of radio transients are known, ranging from extremely nearby
(ultra-high energy cosmic rays impacting the Earth's atmosphere) to
cosmological distances ($\gamma$-ray bursts).  There are also a number 
of classes of hypothesized classes of transients.  In this section we
provide an overview of the wide variety of radio transients that the
SKA could detect and study.

\begin{description}
\item[Ultra-high energy particles:]
Intense, short-duration pulses ($\sim 1$~MJy in $\sim 1$~ns) at
frequencies of a few to a few hundred MHz have been 
observed from the impact of ultra-high energy particles on the Earth's 
atmosphere \cite{w01,fg03,hf03}.  High-energy neutrinos impacting the
lunar regolith should also produce radio pulses near~1~GHz, though searches to
date have been unsuccessful in detecting any such pulses
\cite{heo96}.  Detection of such particles can place significant
constraints on the efficiency of cosmic accelerators, and
potentially on the existence of physics beyond the Standard Model of
particle physics, if particles with energies beyond the
Greisen-Zatsepin-Kuzmin limit ($10^{19.7}$ eV) are detected.

\item[The Sun:]
Type~II and~III solar bursts are detected regularly at radio
frequencies of tens of MHz \cite{mkcasms96,pscdm88} while solar flares
can be detected at tens of GHz \cite{wksysk03}.  It is not yet clear
that the SKA capabilities will extend to frequencies that will
optimize studies of the Sun, but, if so, the Sun offers a nearby site
to study particle acceleration in detail, particularly when
observations are combined with optical, ultraviolet, and X-ray
observations.

\item[Planets:]
Jupiter has long been known to emit radio flares at decameter
wavelengths \cite{abmglr2000,lbrbmk98}, and all of the planets with
strong magnetic fields (Earth, Jupiter, Saturn, Uranus, and Neptune)
produce radio radiation, though not always above the Earth's
ionospheric cutoff.  At least one of the known extrasolar
planets\footnote{%
It is worth noting that the first extrasolar planets were discovered
using radio observations of the pulsar PSR~B1257$+$12, though any
radio emission from these planets is not likely to be detectable.
}
also appears to have a strong magnetic field \cite{swb03}.  By analogy
to the solar system planets, Farrell et al.~\cite{fdz99}, Zarka et
al.~\cite{ztrr01}, and Lazio et al.~\cite{lazioetal04} suggest that
extrasolar planets would produce bursty emission as well.
Characteristic frequencies, based on scaling laws from the solar
system, suggest that the known extrasolar planets would radiate
primarily in the range 10--1000~MHz with flux densities
of~10--100~$\mu$Jy.  Detection of radio emission from extrasolar
planets would constitute \emph{direct} detection, in contrast to the
largely indirect detections of the known extrasolar planets achieved
to date from the reflex motions of their hosts stars.

\item[Brown dwarfs:]
Radio flares have been detected from BD~LP944$-$20
\cite{bergeretal01}, and a survey of late-type stars and brown dwarfs
found a number of other objects also exhibiting flares \cite{b02}.
Typical flare strengths are of order 100~$\mu$Jy at frequencies
between~5 and~8~GHz.  The flares are thought to originate in magnetic
activity on the surfaces of the brown dwarfs, though some objects have 
radio emission that deviates from the expected radio--X-ray
correlation observed for stars.

\item[Flare stars:]
Radio flares from various active stars and star systems are observed
at frequencies of order 1~GHz with flux density levels that can reach
of order 1~Jy \cite{jkw89,rwgr03,g-spr03}
These systems can show strong polarization, including strong circular
polarization.  These flares are attributed to particle acceleration
from magnetic field activity.

\item[Pulsar giant pulses:]
While all pulsars show pulse-to-pulse intensity variations
\cite{hw74}, some pulsars have been found to emit so-called ``giant''
pulses, pulses with strengths 100 or even 1000 times the mean pulse
intensity. The Crab (PSR~B0531$+$21) was the first pulsar found to
exhibit this phenomenon. In one hour of observation, the largest
measured peak pulse flux of the Crab is roughly $\sim$~10$^{5}$~Jy at
430~MHz for a duration of roughly 100~$\mu$s \cite{hr75},
corresponding to an implied brightness temperature of 10$^{31}$
K. Recently, pulses with flux $\sim$ $10^{3}$~Jy at 5~GHz for a
duration of only 2~ns have been detected from the Crab
\cite{hkwe03}. These ``nano-giant'' pulses imply brightness
temperatures of 10$^{38}$~K, by far the most luminous emission from
any astronomical object. For many years, this phenomenon was thought
to be uniquely characteristic of the Crab. However, giant pulses have
since been detected from the millisecond pulsars  PSR~B1821$-$24 \cite{rj01} and
PSR~B1937$+$21 \cite{cstt96} and PSR~B0540$-$69, the Crab-like pulsar in the Large
Magellanic Cloud \cite{jr03}.

\item[Transient pulsars:]
Nice~\cite{n99} searched for radio pulses along~68~deg$^{2}$ of the
Galactic plane at~430~MHz. This search resulted in the detection of
individual pulses from~5 known pulsars and the discovery of one new
pulsar which was previously missed in a standard periodicity search
\cite{nft95}. This pulsar (J1918$+$08) does not emit giant
pulses\footnote{%
We define giant pulses as those comprising a long tail on the overall
pulse amplitude distribution; for the Crab pulsar and the millisecond
pulsars B1937$+$21 and B1821$-$24, these tails are power-law
in form.
}
but is a normal, slow pulsar which fortuitously emitted one strong
pulse during the search observations.  More recently, Kramer et
al.~\cite{krameretal04} have recognized a class of pulsars that produce
pulses only a small fraction of the time.  When ``on,'' they appear
indistinguishable from normal pulsars.  
For instance, the 813-ms pulsar PSR~B1931$+$24 is not detectable more than 90\% of
the time \cite{krameretal04}, and the 1.8-s pulsar PSR~B0826$-$34 is in a
``weak'' mode, thought for many years to be a complete null, roughly 70\%
of the time \cite{esamdinetal04}.
Single-pulse searches of the Parkes Multibeam Pulsar Survey data also have
resulted in the discovery of several pulsars whose emission is so sporadic
that they are not detectable in standard Fourier domain searches \cite{mclaughlinetal04}.

\item[X-ray binaries:]
Large radio flares, with peak flux densities during an outbursts being
factors of~10--100 larger than quiescence, have long been known from
X-binaries such as Cygnus~X-3 \cite{wgjffs95,fbbwpgf97}.  Search for
short-duration, single pulses from the X-ray binaries Scorpius~X-1 and
Cygnus~X-1 have been unsuccessful, though \cite{thh72}.

\item[Soft $\gamma$-ray repeaters:]
Vaughan \& Large~\cite{vl89} conducted an unsuccessful search for
radio pulses from the soft $\gamma$-ray repeater SGR~0526$-$66.

\item[Maser flares:]
The emission from OH masers can vary on timescales of hundreds of
seconds and be detected as long-duration radio bursts \cite{cb85,y86}.

\item[Active galactic nuclei:]
Active galactic nuclei (AGN) outbursts, likely due to propagation of
shocks in relativistic jets, are observed at millimeter and centimeter
wavelengths \cite{aalh85,l94}.

\item[Intraday variability:]
Intraday variability (IDV), resulting from interstellar scintillation
of extremely compact components ($\sim 10$~$\mu$as) in extragalactic
sources, occurs at frequencies near~5~GHz.  The typical modulation
amplitude is a few percent and is both frequency and direction
dependent, but occasional rare sources are seen to display much larger
amplitude modulations with timescales of hours to days \cite{k-cjwtr01,ljbk-cmrt03}.  Intraday
variability could be an important issue for calibration and imaging of
the SKA as existing surveys suggest that it becomes more prevalent at
lower flux densities ($< 100$~mJy).

\item[Radio supernovae:]
With the goal of detecting the single, large, broadband radio pulse
($< 1$~s) expected to
be emitted at the time of a supernova explosion \cite{cn71}, Huguenin
\& Moore~\cite{hm74} and Kardashev et al.~\cite{kardashevetal77}
performed radio searches for single pulses, but found no convincing
signals of extraterrestrial origin aside from solar flares.

\item[$\gamma$-ray bursts:]
In searches for radio pulses associated with $\gamma$-ray bursts,
Cortiglioni et al.~\cite{cmmcis81}, Inzani et al.~\cite{ismm82}, and
Amy et al.~\cite{alv89} detected some dispersed radio pulses, but
found no convincing associations with gamma-ray burst sources.
Balsano~\cite{b99} found a dispersed radio pulse apparently coincident
with GRB980329; however, it was narrowband, which has led to it being
interpreted as due to terrestrial interference.  Various searches for
radio pulses associated with gamma-ray bursts (including precursor
pulses) have been conducted at~151~MHz
\cite{kgwwp94,kgwwp95,desseneetal96}.  Typical upper limits have been 
approximately 100~Jy.  Theoretically, Usov
\& Katz~\cite{uk00} and Sagiv \& Waxman~\cite{sw02} have predicted
that gamma-ray bursts should have associated prompt emission, most
likely below~100~MHz.

\item[Gravitational wave sources:]
In a search for radio counterparts to the gravitational pulses
reported by Weber~\cite{w69}, both Hughes \& Retallack~\cite{hr73} and
Edwards et al.~\cite{ehm74} detected excesses of radio pulses from the
direction of the Galactic center, but did not believe them to be
correlated with the gravitational pulses.  More recently Hansen \&
Lyutikov~\cite{hl01} have predicted that inspiraling neutron
star-neutron star binaries could produce radio precursors to the
expected gravitational wave signature.

\item[Annihilating black holes:]
O'Sullivan et al.~\cite{oes78} and Phinney \& Taylor~\cite{pt79} conducted
searches for radio bursts at frequencies near~1~GHz possibly associated with annihilating black
holes, as suggested by Rees~\cite{r77}, but likewise found no convincing
signals.

\item[Extraterrestrial transmitters:]
While no such examples are known of this class, many searches for
extraterrestrial intelligence (SETI) find non-repeating signals that
are otherwise consistent with the expected signal from an
extraterrestrial transmitter.  Cordes et al.~\cite{cls97} discuss how
extraterrestrial transmitters could appear to be transient, even if
intrinsically steady.
\end{description}

\section{Exploring Phase Space}\label{sec:drs.phasespace}

For definiteness, we consider transients as those objects or emission
phenomena that show substantial flux density changes on time scales of
one month or less.  Although our upper limit on relevant time scales
is arbitary, the lower limit is set by the physics in the source.  For
instance, the pulsar with the shortest known pulse period is
PSR~B1937$+$21 for which $P = 1.56$~ms, while some giant pulses from the
Crab pulsar have substructure at~5~GHz as short as
approximately 2~ns \cite{hkwe03}.  Extensive air showers have been
hypothesized to have time scales as short as 1~ns.

\begin{figure}
\includegraphics[width=\columnwidth]{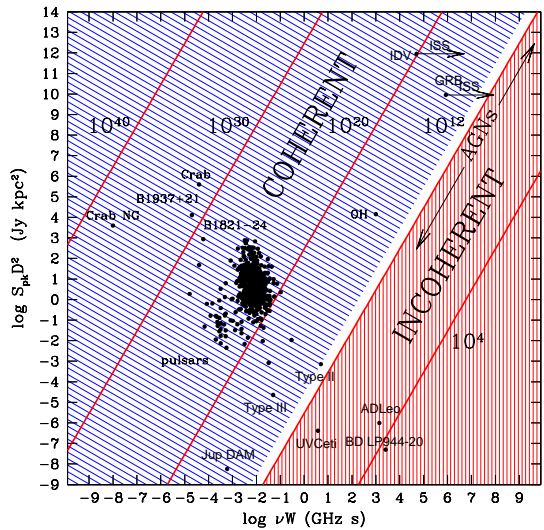}
\vspace{-1.5cm}
\caption[]{The phase space for radio transients.  The abscissa is the
product of the emission frequency $\nu$ and transient duration or pulse
width~$W$.  The ordinate is the product of the observed flux
density~$S$ and square of the distance~$D^2$.  In the Rayleigh-Jeans approximation,
these quantities are directly proportional and related to the
brightness temperature~$T$ (eqn.~\ref{eqn:drs.phasespace}).  The sloping
lines are labelled by constant brightness temperature.  A brightness
temperature of~$10^{12}$~K is taken to divide coherent from incoherent 
sources.  Examples of transient emission from various classes of
sources are indicated.  In the case of gamma-ray burst afterglows
(GRB) and intraday variability (IDV) of active galactic nuclei (AGN),
the apparently high brightness temperatures are not thought to be
intrinsic but related to interstellar scintillation (ISS).  For these
two classes of sources we show how the absence of ISS would affect
their locations in this phase space.}
\label{fig:drs.phasespace}
\end{figure}

In the Rayleigh-Jeans approximation, a source with brightness
temperature~$T$ varies (intrinsically) on a time scale or pulse
width~$W$ given by
\begin{equation}
W^2 = \frac{1}{2\pi k}\frac{SD^2}{T}\frac{1}{\nu^2},
\label{eqn:drs.phasespace}
\end{equation}
where the observed flux density is $S$, the source's distance is $D$,
the emission frequency is $\nu$, and $k$ is Boltzmann's constant.
Figure~\ref{fig:drs.phasespace} shows the phase space of the
pseudo-luminosity $SD^2$ vs.\ $\nu W$.  There are two notable aspects
of this figure.  First, the transient radio sources observed span a
large range in this phase space.  The range of $\nu W$ covers at least
13 orders of magnitude while the range of $SD^2$ covers at least 20
orders of magnitude.  Second, large portions of this phase space are
empty.

One of two conclusions can be drawn from
Figure~\ref{fig:drs.phasespace}.  One could conclude no physical
mechanisms or sources exist that would populate the empty regions in
the phase space of Figure~\ref{fig:drs.phasespace}.  Alternately, we
regard Figure~\ref{fig:drs.phasespace} as an illustration of the
incompleteness of our knowledge of the transient radio sky and of the
potential for the \hbox{SKA}.

\begin{figure}
\includegraphics[width=\columnwidth]{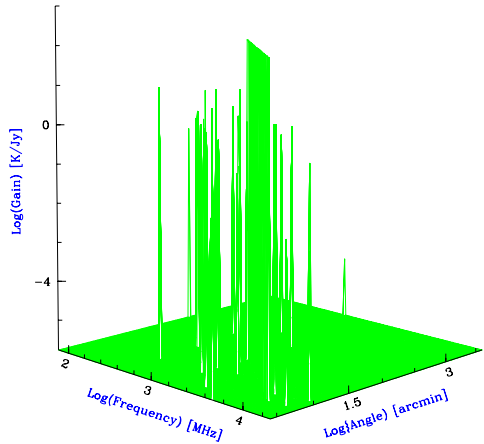}
\vspace{-2cm}
\caption[]{A representation of the combined frequency coverage, solid
angle coverage, and sensitivity of a variety of historical and
existing radio telescopes.  The solid angle coverage is shown as the
FWHM dimension of the beam (or equivalent) in arcminutes.  The
sensitivity is shown as the gain in units of~K~Jy${}^{-1}$.  All
scales are \emph{logarithmic}.  For reference, the largest angle
coverage is provided by the 610~MHz STARE survey with a FWHM beam
equivalent of~$4100^\prime$ while the most sensitive telescope other
than the SKA is the
Arecibo telescope with a 430~MHz gain of~11~K~Jy${}^{-1}$.  The SKA is 
shown assuming that it has a $60^\prime$ field of view at~1~GHz, which 
scales as $\nu^{-1}$, and that it has an effective collecting area in
the ``core'' of 500\,000~m${}^2$, which is constant between~1
and~10~GHz.  Not shown is the EVLA which will have a similar frequency 
coverage but an angle coverage of~4 times smaller and a sensitivity
of~75 times less.}
\label{fig:drs.telescopes}
\end{figure}

In order for the SKA to explore the dynamic radio sky,
``equation~(\ref{eqn:drs.fom})'' must be satisfied; it must obtain
large solid angle coverage, high time resolution, and high sensitivity
simultaneously.  We discuss the time resolution requirements in more
detail below.  Figure~\ref{fig:drs.telescopes} illustrates that,
historically, radio telescopes have been capable of obtaining either
large solid angle coverage (e.g., the STARE survey at~610~MHz by Katz
et al.~\cite{khcm03} with a FWHM beam of~$4000^\prime$) or high
sensitivity (e.g., the Arecibo telescope with a gain
of~11~K~Jy${}^{-1}$ at~430~MHz) but not the two simultaneously.  If
the SKA satisfies its design requirements, it will produce at least an
order of magntiude sensitivity improvement, over a large frequency
range, with a solid angle coverage that is comparable to or exceeds
that of all but a small number of low-sensitivity telescopes.

\section{SKA Operational Modes and Experiments}\label{sec:drs.survey}

The SKA is most likely to open up new areas of parameter space or
discover new classes of sources by conducting dedicated transient
searches.  For a fixed collecting area~$A$ and fixed total telescope
time, the transient figure of merit (equation~\ref{eqn:drs.fom}) can
be optimized either by covering as much solid angle as possible
(``tiling'') or by observing as deeply as possible (``staring'').
Various optimizations are possible, depending upon the class of
transient being sought, what is known about the typical duration and
sky distribution of the class, and the importance of interstellar
scintillations \cite{cl91,n03,cm03}.  We consider a spectrum of
possible transient searches, ranging from a ``pure'' staring search
(extrasolar planets) to a ``pure'' tiling search (an all-sky survey).
Our examples are by no means the only classes of potential transients
that are amenable to these methods of searching, but they represent
classes for which the sensitivity of the SKA is either essential or
produces qualitatively larger numbers of such transients to be
detected.

In addition, we envision that a host of traditional radio transient
studies will be possible with the \hbox{SKA}.  Much like current
interferometers, the SKA will be able to respond to triggers from
other wavelengths.  Current examples of this capability include
responding to $\gamma$-ray bursts, supernovae, and X-ray transients.
With its high sensitivity and high angular resolution, the SKA will
both continue and improve on existing studies.

The SKA should also be capable of conducting monitoring experiments,
possibly used to trigger observations at other wavelengths.  In many
current concepts, the SKA is envisioned as having a ``core'' and
``outlying stations.''  (Indeed, below we discuss why such a
configuration is necessary for the SKA for transient searching.)  Not
all observations will be able to make use of the entire SKA, e.g., use
of the outlying stations produces too low of a surface brightness
sensitivity for some kinds of H\,\textsc{i} observations.  Thus, a
small number of outlying stations could be used to form a subarray and
tasked to measure the flux density of a variety of objects, such as to
search for changes resulting from flares from X-ray binaries or
extreme scattering events (ESEs) toward active galactic nuclei.

\subsection{Staring at Extrasolar Planets}\label{sec:drs.exoplanet}

The past few years have been an exciting time as extrasolar planets
have been demonstrated to be widespread and multiple planetary systems
have been found.  The current census now numbers more than 100
extrasolar planets, in over 90 planetary systems \cite{s03,m04}.

The vast majority of these extrasolar planets have been detected via
the reflex motion of the host star.  As the existing census shows,
this method has proven to be wildly successful.  Nonetheless, the
reflex motion of the star is a measure of the planet's gravitational
influence and is necessarily an \emph{indirect} detection of the
planet.  As a consequence, the only property of the planet that one
can infer is its mass, and because of the mass function's dependence
on the inclination angle ($\sin i$), one can infer only a minimum
mass.

Direct detection of reflected, absorbed, or emitted radiation from a
planet allows for the possibility of complementary information, and
likely a more complete characterization of the planet.  The prototype
of such a direct detection is the detection of sodium absorption lines
in the atmosphere of the planet orbiting HD~209458 \cite{cbng02}.
Unfortunately, the incidence of transiting planets will always remain
low relative to the total number of planets known.

The Earth and gas giants of our solar system are ``magnetic planets''
in which internal dynamo currents generate a planetary-scale magnetic
field.  The radio emission from these planets arises from coherent
cyclotron emission due to energetic (keV) electrons propagating along
magnetic field lines into active auroral regions \cite{g74,wl79}.  The
source of the electron acceleration to high energies is ultimately a
coupling between the incident solar wind and the planet's magnetic
field, presumably due to magnetic field reconnection in which the
magnetic field embedded in the solar wind and the planetary magnetic
field cancel at their interface, thereby energizing the plasma.
Energetic electrons in the energized plasma form a current flow
planet-ward along the planet's magnetic field lines, with the lines
acting effectively like low resistance wires.  The energy in these
magnetic field-aligned electric currents is deposited in the upper
polar atmosphere and is responsible for the visible aurora.  Besides
auroral emissions at visible wavelengths (e.g., northern lights),
about~$10^{-5}$--$10^{-6}$ of the solar wind input power is converted
to escaping cyclotron radio emission \cite{g74}.  The auroral radio
power from Jupiter is of order $10^{10.5}$~\hbox{W}.

At least one of the known extrasolar planets is also a magnetic
planet.  Shkolnik, Walker, \& Bohlender~\cite{swb03} have detected a
modulation in the Ca~\textsc{ii}~H and~K lines of HD~179949 with a
periodicity which is that of the planetary orbit.  They interpret this
as a magnetic interaction between the star and planet, though there is
no constraint as yet on the magnetic field strength of the planet.

The SKA offers the possibility of detecting radio emission from
extrasolar planets.  Detection of radio emission from an extrasolar
planet would constitute a \emph{direct} detection and can yield
fundamental information about the planet.  First, a measurement of the
radio emission is directly indicative of the polar magnetic field
strength at the planet.  For example, the high-frequency cutoff of
Jovian decametric bursts ($\simeq 40$~MHz) is interpreted as being due
to the Jovian polar magnetic field strength, which allowed an estimate
for the strength of the Jovian magnetic field nearly~20~yr prior to
the first \textit{in situ} magnetic field observations.  In turn, the
presence of a magnetic field provides a rough measure of the
composition of the planet, insofar as it requires the planet's
interior to have a conducting fluid.  Combined with an estimate of the
planet's mass, one could deduce the composition of the fluid by
analogy to the solar system planets (liquid iron vs.\ metallic
hydrogen vs.\ salty ocean).

Second, the periodic nature of the radio emission has been used to
define precisely the planetary rotation periods of all of the gas
giant planets in the solar system, because the magnetic field is
presumed to be tied to the interior of the planet.

Finally, testing the extent to which solar-system models of magnetic
fields can be applied to extrasolar planets may have important
implications for assessing the long-term ``habitability'' of
terrestrial planets found in the future.  The importance of a magnetic
field is that it deflects incident cosmic rays.  If these particles
reach the surface of an otherwise habitable planet, they may cause
severe cellular damage and disruption of genetic material to any life
on its surface or may prevent life from arising at all.  A secondary
importance of a magnetic field is that it can prevent the planet's
atmosphere from being eroded by the stellar wind
\cite{mitchelletal01}; this process is thought to be a contributing
factor to the relative thinness of the Martian atmosphere.  (This is
of course unlikely to be an issue for extrasolar giant planets.)

Farrell et al.~\cite{fdz99}, Zarka et al.~\cite{ztrr01}, and Lazio et
al.~\cite{lazioetal04}, building on empirical relations derived for
solar system planets and calibrated by spacecraft fly-bys, have made
specific predictions for the radio emission from the known extrasolar
planets.  Their predictions have made use of two empirical relations,
\emph{Blackett's Law} and the \emph{Radiometric Bode's Law}.

Blackett's law relates a planet's magnetic moment (its surface field
times its radius cubed, $BR^3$) to its rotation rate and mass.  The
Radiometric Bode's law relates the incident stellar wind power and the
planet's magnetic field strength to its median emitted radio power.

The predicted radio power for a magnetic planet immersed in its host's
star stellar wind is
\begin{eqnarray}
P_{\mathrm{rad}} &\sim& 4 \times 10^{11}\,\mathrm{W}\nonumber\\
 & &\left(\frac{\omega}{10\,\mathrm{hr}}\right)^{0.79}\left(\frac{M}{M_J}\right)^{1.33}\left(\frac{d}{5\,\mathrm{AU}}\right)^{-1.6},\nonumber\\
\label{eqn:rbl}
\end{eqnarray}
where $\omega$ is the planet's rotation rate, $M$ is its mass, and $d$ 
is the distance between the planet and host star.  All quantities have been normalized to those of Jupiter.
Farrell et al.~\cite{fdz99} discussed slight differences to this radiometric Bode's
law derived by various authors; the differences result from the
statistical spread in the various (solar system) planets' magnetic
moments and amount to slightly different exponents and/or a different
coefficient.

We stress various aspects of this radiometric Bode's law.  First, it
is grounded in \textit{in situ} measurements from spacecraft fly-bys
of the gas giants as well as measurements of the Earth's radio
emission.

Second is the importance of the planet-star distance.  As
Zarka et al.~\cite{ztrr01} show, the radio power from Earth is larger than that
from Uranus or Neptune even though both of those planets have magnetic
moments approximately 50~times larger than that of the Earth.

Third, the radiometric Bode's Law of equation~(\ref{eqn:rbl})
describes the \emph{median} emitted power from the magnetospheres of
the Earth and all of the solar gas giants, including the \emph{non-Io
driven} Jovian decametric radio emission \cite{ztrr01}.  Planetary
magnetospheres tend to act as ``amplifiers'' of the incident solar
wind, so that an increase in the solar wind velocity (and therefore
incident pressure) leads to geometrically higher emission levels.
This effect is observed at all of the magnetized planets.  Based on the
range of solar wind velocities and emitted radio powers observed at
the Earth \cite[Figure~5][e.g.,]{gd81}, the radio power levels from a
planet can exceed that predicted by equation~(\ref{eqn:rbl}) by
factors of~100 and possibly more.

\begin{figure}
\includegraphics[angle=-90,width=\columnwidth]{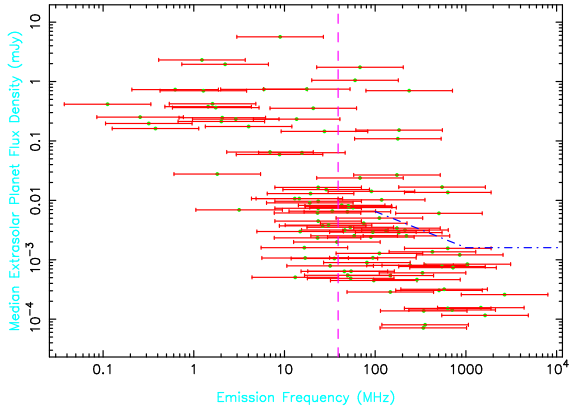}
\vspace{-1cm}
\caption[]{The \emph{median} predicted flux densities for 106 known
extrasolar planets vs.\ the characteristic emission frequency based on
the radiometric Bode's Law and Blackett's Law.  The horizontal bars
indicate the assumed ranges for the emission frequencies, allowing the
statistical variations from Blackett's Law in the estimated planetary
magnetic moments.  The vertical dashed line indicates the 
cutoff frequency for Jupiter's radio emission.  The dot-dashed line shows the
sensitivity of the SKA, assuming an integration time of~15~min.\ and a
bandwidth of~4~MHz.  For the SKA, a nominal sensitivity of
$A_{\mathrm{eff}}/T_{\mathrm{sys}} =
5000\,\mathrm{m}^2\,\mathrm{K}^{-1}$ has been assumed at~100~MHz,
improving to $20\,000\,\mathrm{m}^2\,\mathrm{K}^{-1}$ at~1000~MHz.}
\label{fig:census}
\end{figure}

The required quantities for applying equation~(\ref{eqn:rbl}) are the
planet's mass, distance from its primary, and rotation rate.  The
radial velocity method determines a lower limit to the planet's mass
(i.e., $M\sin i$) and the semi-major axis of its orbit.  We have no
information on the rotation rates of these planets.  We have therefore
assumed that those planets with semi-major axes larger than 0.1~AU
have rotation rates equal to that of Jupiter (10~hr), while those with
semi-major axes less than 0.1~AU are tidally-locked with rotation
rates equal to their orbital periods \cite{gs66,t00}.
Figure~\ref{fig:census} presents the expected \emph{median} flux
densities vs.\ the emission frequency in a graphical form.

The trend in Figure~\ref{fig:census} of increasing emission frequency
and decreasing flux density is real and reflects two effects.  First,
the lower envelope reflects a selection effect.  A low flux density
and small emission frequency results from a low-mass planet in a large
orbit.  These planets cannot be detected with the current detection
methodology.  Second, the upper envelope reflects the well-known
deficit of planets with both large masses and small semi-major axes.
Even if high-mass planets with close orbits did exist, however, they
probably would be tidally locked (as we have assumed here).  The
rotation rate also determines the strength of the magnetic field, so
tidally-locked planets probably do not radiate at high emission
frequencies.  However, \cite{z04} have suggested that such ``hot
Jupiters'' may radiate by the conversion of stellar magnetic pressure
(which is large in close to the star) into electromotive forces,
currents, and radio emission.

In designing an experiment to detect extrasolar planets, staring is
clearly the preferred method as the locations of the planets are known
and covering large solid angles would not increase the odds of
detection.  However, the potential for bursts must be taken into
account.  The sensitivity of a radio telescope improves with
increasing integration time as $t^{-1/2}$.  Longer integration times
improve the sensitivity but at the cost of ``diluting'' any bursts.
For a burst of flux density~$S_b$ and duration~$\Delta t_b$, its
average flux density in an integration time~$t$ is $S_b(\Delta
t_b/t)$, i.e., the burst is diluted with time as $t^{-1}$.  To the
extent that extrasolar planet radio emission is ``bursty'' (as is
observed for the solar system planets), multiple short observations
are a better strategy than a single long integration.

Current design goals for the SKA specify its sensitivity to be
$A_{\mathrm{eff}}/T_{\mathrm{sys}} =
5000\,\mathrm{m}^2\,\mathrm{K}^{-1}$ at~100~MHz, improving to
$20\,000\,\mathrm{m}^2\,\mathrm{K}^{-1}$ at~1000~MHz.  In a 15~min.\
integration, its flux density sensitivity would be in the range
1--10~$\mu$Jy, more than sufficient to detect the radio emissions from
the most massive extrasolar planets, without relying upon bursts to
enhance the emission levels.  If bursts are considered, the SKA should
detect a substantial fraction of the current census, if the bursts are
comparable in magnitude to what is seen in the solar system (factors
of~10--100).

\subsection{Giant Pulses from Extragalactic Pulsars}\label{sec:drs.gp}

As discussed above, the Crab pulsar and a small number of other
pulsars are known to emit giant pulses, pulses that comprise a long
tail on the overall pulse amplitude distribution.  These probably
arise from coherent emission within the turbulent plasma in the pulsar
magnetosphere \cite{hkwe03}.  Some of these pulses are so strong that,
were the Crab pulsar located in a nearby external galaxy, it could be
detected with existing telescopes.  Exploiting both the SKA's sheer
sensitivity and flexibility, giant pulses from pulsars in external
galaxies should be detectable well beyond the Local Group and
potentially to the distance of the Virgo Cluster.

The fraction of its lifetime over which a pulsar might exhibit the
giant pulse phenomenon is not known.  If we take the Crab as an
exemplar of young, giant-pulse emitting pulsars, we might expect a
birth rate of order 1 pulsar per 100~yr and that the giant pulse
phenomenon lasts for of order 1000~yr.  With careful accounting for
selection effects (e.g., pulsar beaming fraction), giant-pulse
emitting pulsars in external galaxies can probe the recent massive
star formation in those galaxies.

Moreover, by their dispersion measures (DM) and rotation measures (RM),
extragalactic pulsars have the potential to probe the baryon density
and magnetic field in the Local Group and potentially beyond.  Current
models for large-scale structure formation predict that matter in the
current epoch forms a ``cosmic web'' \cite{co99,dkhw99,daveetal01}.
Most of the baryons in the Universe reside in large-scale filaments
that form the ``strands'' of the web, with groups and clusters of
galaxies located at the intersections of filaments.  At the current
epoch, hydrogen gas continues to accrete onto and stream along these
filaments, undergoing multiple shocks as it falls into the
gravitational potential wells located at the intersections of the
filaments.  In this model, the filaments of hydrogen form a warm-hot
ionized medium (WHIM), with a temperature of~$10^5$--$10^7$~\hbox{K}.

Recent observations of highly ionized species of oxygen and neon by
both the \emph{FUSE} and \emph{Chandra} observatories are considered
to be validations of these predictions.  Absorption observations along
various lines of sight suggest a diffuse medium with a temperature of
order $10^6$~K with a density of order $5 \times 10^{-5}$~cm${}^{-3}$.
While striking, these observations still suffer from the difficulty of
probing only trace elements.  The ionized hydrogen in the WHIM has not
been detected directly, but, over megaparsec path lengths, the implied
DMs are of order 50~pc~cm${}^{-3}$, comparable to or larger than that
for nearby pulsars.

\begin{figure}
\vspace*{-1cm}
\includegraphics[angle=-90,width=\columnwidth]{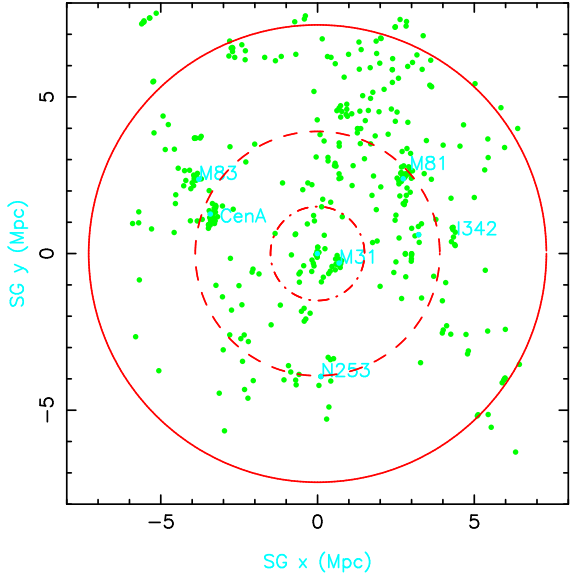}
\vspace{-1cm}
\caption[]{The distance to which a Crab giant pulse can be detected.
The three circles show (in order of increasing size) the distance to
which the strongest pulse detected in~1~hr using Arecibo (5\% of the
\hbox{SKA}, assuming that it could see the entire sky), 30\% of the
\hbox{SKA}, and the full \hbox{SKA}.  The distribution of local
galaxies is shown in the Supergalactic $x$ and~$y$ planes (from
\cite{kkhm04}), and certain galaxies or groups of galaxies are labelled.}
\label{fig:drs.giantpulse}
\end{figure}

For the Crab pulsar, the most well-studied giant-pulse emitting
pulsar, the giant pulse amplitude distribution is power-law in form at
high amplitudes \cite{lcumlfh95,cbhmk03}.  On average, the strongest
pulse observed from the Crab pulsar at~0.43~GHz in one hour has a
signal-to-noise ratio of ${S/N}_{\rm max} \approx 10^4$, even with the
system noise dominated by the Crab Nebula.  For objects in other
galaxies, the system noise is essentially unaffected by any potential
nebular contribution.  If the Crab pulsar were not embedded in its
nebula, the $S/N$ of such a pulse would be larger by the ratio of the
system noise equivalent flux density when observing the Crab Nebula to the
nominal system noise equivalent flux density or $S_{\rm CN} / \Ssyso
\approx 300$ times larger, or $3.3\times 10^6$.  Thus, on average, in
an hour's observation, one should be able to detect a Crab-like pulsar
to a maximum distance at a specified signal-to-noise ratio,
$(S/N)_{\rm det}$,
\begin{eqnarray}
&D_{\rm max}& \nonumber\\
        &=& \DCN
        \left [
                \frac {(S/N)_{\rm max}} { (S/N)_{\rm det}}
                \left ( 1 + \frac{\SCN}{\Ssyso} \right)
                - \frac{\SCN}{\Ssyso}
        \right ]^{1/2} \nonumber\\
        &\approx&  1.6\,{\rm Mpc}
        \left[\frac{(S/N)_{\rm det}}{5}\right] ^{-1/2}
        \left (\frac{f_c A_{\rm SKA}}{A_{\rm AO}} \right)^{1/2},\nonumber\\
\label{eqn:drs.gpmax}
\end{eqnarray}
where $f_c A_{\rm SKA}$ is the SKA's collecting area that can be used
for a giant pulse survey and $A_{\rm AO}$ is Arecibo's effective area
at~0.43~GHz.  The usable area for the SKA in a blind survey will be
limited by what fraction~$f_c$ of the antennas are directly connected
to a central correlator/beamformer.  For $f_c = 1$, $A_{\rm SKA} /
A_{\rm AO} \approx 20$ so the standard one-hour pulse seen at Arecibo
could be detected to~7.3~Mpc.  Figure~\ref{fig:drs.giantpulse} shows
that even for more modest values of $f_c$, e.g., $f_c \approx 0.3$,
the SKA could detect giant-pulse emitting pulsars well beyond the
Local Group.

No upper amplitude cutoff has been observed yet for the Crab pulsar's
giant pulse amplitude distribution \cite{lcumlfh95}.  Observing for
longer than 1~hr results in even stronger pulses being detected.
Assuming that the Crab pulsar is not atypical, a modest extrapolation
of the giant pulse amplitude distribution implies that the SKA could
detect giant-pulse emitting pulsars in the Virgo cluster ($D \approx
20$~Mpc).

An additional efficiency is obtained toward the Virgo cluster as many
galaxies will be contained within the SKA's field of view.  Although
it is not yet clear what frequency would be optimal for a giant-pulse
search toward the Virgo cluster, it almost certainly will be not much
higher than 1~GHz.  Thus, the field of view will be at least
1~deg.${}^2$ and could be potentially nearly 10 times larger, e.g.,
for observations at~0.33~GHz.

The baryon density along the line of sight to an extragalactic
giant-pulse emitting pulsar can be inferred from the pulsar's
\hbox{DM}.  The dispersion measure is simply the line-of-sight
integral of the electron density, $\mathrm{DM} = \int n_e\,dl$.  As is
the case for more traditional pulsar searches, giant pulse searches
require dedispersion, which is accomplished by searching through a
number of trial DMs \cite{mc03}.  As a line-of-sight integral, the DM
for an extragalactic pulsar contains contributions from the Galaxy,
the intergalactic medium, and the host galaxy(ies).  The Galactic
contribution to the DM can be predicted from models for the Galactic
free electron distribution \cite{cl02}, and in the SKA era even more
detailed models will be available.  However, the contribution from the
host galaxy will be able to be estimated only crudely (e.g., from its
inclination and assuming a plane-parallel electron density
distribution).  We anticipate that the uncertainty in separating the
host galaxy and intergalactic contributions to the DMs of
extragalactic pulsars will scale roughly as $\sqrt{N}$ for $N$
extragalactic pulsars.  Thus, it is not sufficient merely to detect
one or a few extragalactic pulsars, as might be accomplished with
existing telescopes (like the 100-m Efflesberg telescope, the Green
Bank Telescope, or Arecibo).

Similar considerations apply to the measurement of the intergalactic
magnetic field via the rotation measure, $\mathrm{RM} = \int n_e
\mathbf{B}\cdot\mathbf{dl}$.  Measurement of the RM for extragalactic
pulsars would prove a unique probe of the magnetic field in the Local
Supercluster, but this requires the detection of many extragalactic
pulsars to separate the Galactic, host galaxy, and intergalactic
contributions.

Searching for giant pulses from extragalactic pulsars requires
time-domain searches, i.e., it is a \emph{non-imaging} application of
the \hbox{SKA}.  In
order to conduct these searches, there are three key requirements for
the \hbox{SKA}.  The first is that it have the capability to provide,
in a routine manner, fast-sampled data.  Given that part of the
motivation for transient searching is to open up new parameter space,
an arbitarily fast sampling may appear necessary.  There are, however, 
certain physically-motivated limits that can be specified.  First, in
order to accomplish the Key Science Driver of ``Strong Field Tests of
Gravity Using Pulsars and Black Holes'' a minimum time sampling
of~10~$\mu$s is required in order to conduct pulsar timing
observations.  Lazio \& Cordes~\cite{lc03} summarized various
interstellar propagation effects that suggest that a time sampling
faster than a few microseconds is not justified, at least for sources
at interstellar or intergalactic distances at frequencies near~1~GHz.
These effects are strongly frequency dependent, though, so nanosecond
time sampling can be tolerated at frequencies above~5~GHz, as
evidenced by the Crab pulsar's nano-giant pulses.  Finally, the
coherent pulses from ultra-high energy cosmic ray impacts on the
Earth's atmosphere have time scales of order 1~ns.  Thus, a useful
target range for the SKA would be sampling times between~1~ns
and~10~$\mu$s.

The second requirement for the SKA is on its configuration.
Backer~\cite{dcb99} has shown that the signal-to-noise ratio achieved
in a search scales with the array filling factor~$f$ as $\sqrt{f}$.
Thus, the SKA should have a reasonably compact ``core'' where the
filling factor is not too small.  A useful goal might be a core
containing a significant fraction of the SKA collecting area ($\sim
50$\%) with a filling factor of a few percent to 10\%.  

A third requirement is on effective suppression of radio frequency
interference (RFI).  This requirement is not unique to searches for
transients, of course, but it is perhaps more stringent for transient
searching than for other operational modes.  One wishes to ensure that
cosmic transient signals are not confused with terrestrial RFI and
thereby eliminated.  A variety of techniques may end up being applied
with the \hbox{SKA}, but two key techniques for extragalactic giant
pulses and other fast transients are utilizing the spatial extent of
the SKA and the expected dispersion smearing of the signals.  Cordes
\& McLaughlin~\cite{cm03} illustrate how dispersion smearing can be
used advantageously.  Briefly, any pulse from a source outside the
solar system will suffer dispersion smearing while most terrestrial
RFI will not.  Similarly, terrestrial RFI should not affect all SKA
antennas equally whereas a celestial pulse should, taking into account
the time delay between the antennas.  The spatial extent of the SKA
(or of its core) can be used as an anti-coicidence filter to identify \hbox{RFI}.

Finally, targeted transient searches could be conducted in a
``piggyback'' synthesis mode.  The relatively large field of view of
the SKA implies that most continuum synthesis observations will be
conducted in a pseudo-continuum mode, so as to be able to image the
entire field of view without significant bandwidth smearing.  The
positions and flux densities of most of the ``strong'' sources ($>
10$~mJy) in any given field of view should be known reasonably
accurately (from surveys like the NVSS) and could be subtracted from
the visibility data.  The residual visibility data could then be
imaged on different time scales to search for weak point sources that
vary.  Such a piggyback mode would be particularly effective at
finding moderate duration transients in nearby galaxies and clusters
of galaxies.

\subsection{Tiling the Sky: An All-Sky Survey}\label{sec:drs.allsky}

An all-sky survey would be designed to search for transients of
unknown or poorly known duration and distribution.  As a strawman for
such a survey, we take the goal of surveying an entire hemisphere
within~24~hr.  If the SKA meets its current specification of a
1~deg${}^2$ field of view, then each pointing can be 1~s in duration
(assuming Nyquist sampling of the sky).  Allowing for 50\% efficiency,
assuming a bandwidth of only 1~MHz, and only half of the collecting
area of the SKA (in its core), such a blind, all-sky survey would
still obtain a typical noise level of approximately 0.5~mJy.  Larger
bandwidths or a larger fraction of the SKA collecting area in the core
or both would improve upon this noise level, though larger bandwidths
could also imply a heavier computational load in order to cope with
increased DM searching.  Similarly, the sky could be covered more
quickly at lower sensitivity by dividing the SKA into sub-arrays, each
having the same 1~deg.${}^2$ field of view but reduced sensitivity.

One manner in which an all-sky search would not be challenging would
be the slew rate.  The implied slew rate is comparable to what current 
instruments provide, e.g., for the VLA the slew rate is approximately
0.5~deg.~s${}^{-1}$.

A less ambitious ``blind'' survey would be a survey along the Galactic 
plane.  Existing Galactic plane surveys \cite{lmdekz00} have provided
intriguing hints of transients to be detected.  Such a survey might
also be possible in the early years of the SKA, before the full
computational infrastructure is in place or as an initial effort
before searching the entire sky.  As a strawman survey we consider a
12-hr survey covering $90^\circ$ of Galactic longitude and
$10^\circ$ in Galactic latitude ($|b| < 5^\circ$).  Covering
900~deg${}^2$, again with a 1~deg${}^2$ field of view at~1~GHz, would
allow for approximately 10~s per pointing and produce an rms noise of
approximately 0.1~mJy.  Alternately, a higher frequency could be used
(to reduce the impact of dispersion smearing or any possible pulse
broadening) at the cost of less time per pointing, a slightly
higher noise level, and a loss of sensitivity to steep spectrum sources.  For instance, at~3~GHz, the time per pointing
would be 1~s.

A key aspect of transient searches will be sufficient computational
power.  A full analysis will not be possible until the SKA concept is
chosen (as that will affect issues such as the field of view).  For a
strawman analysis, though, consider an imaging-based search in which a 
correlator produces an image which is then examined for transients.
Preliminary considerations imply that imaging the full FOV is more
efficiently done through correlation than through direct beam forming.

At each integration time the number of correlations to be computed
is 
\be
N_{\rm c} = \half \na(\na-1) \Npol\Nnu,
\ee
where $\na$ is the number of antennas, $\Npol$ is the number of polarization
channels ($\Npol = 4$ for full Stokes), and $\Nnu$ is the number of
frequency channels. 
The number of pixels in the FOV  for a maximum baseline, $\bc$, 
in the core array and an aperture diameter $D$ is 
\begin{eqnarray}
\Npix &\approx& 0.85\left( \frac{\bc}{D}\right)^2 \nonumber\\
      &\approx& 10^4\,\mathrm{pixels} \left ( \bckm / D_{10} \right)^2.
\end{eqnarray}
For fast transients (e.g. durations $\lesssim 1$ sec), 
sufficient channels are needed to allow dedispersion,
a requirement that also satisfies the need for the ``delay-beam'' to 
be large enough for full FOV mapping. Dedispersion in blind surveys
requires summing over frequency with trial values of DM, whose number
is approximately the same as the number of channels. Additional processing
would include matched filtering to identify individual transients
and Fourier analysis to find periodic sources.  For SETI and other
spectral-domain searches, each pixel would be Fourier analyzed
(before detection) to identify candidate spectral lines.

\section{Conclusions}\label{sec:drs.conclude}

Transient radio sources offer insight into dynamic or explosive events
as well as being powerful probes of intervening media.  
Historically, the radio sky also has been searched only poorly for radio
transients, meaning that the SKA has the potential to explore
previously-unexplored parameter space at radio wavelengths.

In order to be effective at searching for transient radio sources, the
SKA must achive a ``large'' value for the figure of merit
$A\Omega(T/\Delta t)$, equation~(\ref{eqn:drs.fom}).  By its very
nature, $A$ is anticipated to be large for the
\hbox{SKA}.  Table~\ref{tab:drs.survey} summarizes the other
requirements on the \hbox{SKA}.  These requirements flow either from
the desire to achieve a large figure of merit or are based on
experience from known radio transients.  

\begin{table}
\caption{SKA Requirements for Transient	Surveys\label{tab:drs.survey}}
\begin{center}
\begin{tabular}{lc}
\noalign{\hrule}
Parameter & Requirement \\
\noalign{\hrule\hrule}
Field of View (at~1~GHz) & 1~deg.${}^2$ \\
Time Sampling            & $\sim 1$~ns to 10~$\mu$s \\
Frequency Coverage       & $\sim 0.1$--10~GHz \\
Configuration            & ``large'' core filling factor \\
                         & long baselines \\
\noalign{\hrule}
\end{tabular}
\end{center}
\end{table}

In order to conduct successfully the Key Science Project, ``Strong
Field Tests of Gravity Using Pulsars and Black Holes,'' the SKA is
required to obtain high time resolution.  The challenge for the
\hbox{SKA}, and historically the difficulty with radio telescopes, is
that they have not provided a large field of view simultaneously with
high sensitivity.  A related requirement is on the configuration of
the array, as the filling factor of the array affects the
signal-to-noise ratio in a search as $\sqrt{f}$.  Thus, the SKA
requires a core, containing a significant fraction of the collecting
area with a reasonablely high filling factor, as well as long
baselines, for localizing sources and imaging searches such as for
$\gamma$-ray bursts and extrasolar planets.

From known classes of sources alone, this search promises to be
fruitful---one can explore particle acceleration by observing radio
pulses from ultra-high energy particles impacting the Earth's
atmosphere as well as from flares from the Sun, brown dwarfs, flare
stars and X-ray binaries while using afterglows from $\gamma$-ray bursts one can
explore the cosmological star formation history.  Likely, though not
yet detected, classes of sources include giant pulses from
extragalactic pulsars, which would allow a direct probe of the local
intergalactic medium, and extrasolar planets, which would be a direct
detection of these objects.

Most exciting would be the \emph{discovery of new classes of sources}.


\begin{thebibliography}{999}

\bibitem{aalh85}
	Aller, H.~D., Aller, M.~F., Latimer, G.~E., \& Hodge, P.~E.
	1985, ApJS, 59, 513

\bibitem{alv89}
	Amy, S.~W., Large, M.~I., \& Vaughan, A.~E.\ 1989,
	Proc.\ Astron.\ Soc.\ Australia, 8, 172

\bibitem{abmglr2000}
	Aubier, A., Boudjada, M.~Y., Moreau, P., Galopeau, P.~H.~M.,
	Lecacheux, A., \& Rucker, H.~O.  2000, A\&A, 354, 1101

\bibitem{dcb99}
	Backer, D.~C.  1999, in Perspectives on Radio Astronomy:
	Science with Large Antenna Arrays, ed.\ M.~P.~van~Haarlem
	(ASTRON: Dwingeloo, The Netherlands) p.~285

\bibitem{b99}
	Balsano, R.~J.  1999, \hbox{Ph.}D.\ Thesis, Princeton U.

\bibitem{b02}
	Berger, E.  2002, ApJ, 572, 503

\bibitem{bergeretal01} Berger, E., et al.  2001, Nature, 410, 338

\bibitem{co99}
	Cen, R.\ \& Ostriker, J.~P.  1999, ApJ, 514, 1

\bibitem{cbng02}
	Charbonneau, D., Brown, T.~M., Noyes, R.~W., \& Gilliland,
	R.~L.  2002, 
	ApJ, 568, 377

\bibitem{cstt96}
	Cognard, I., Shrauner, J.~A., Taylor, J.~H., \& Thorsett,
	S.~E.  1996, ApJ, 457, L81

\bibitem{cb85} Cohen, R.~J.~\& Brebner, G.~C.  1985, MNRAS, 216, 51P

\bibitem{cn71}
	Colgate, S.~A.\ \& Noerdlinger, P.~D.  1971, ApJ, 165, 509

\bibitem{cbhmk03}
	Cordes, J.~M., Bhat, N.~D.~R., Hankins, T.~H., McLaughlin,
	M.~A., \& Kern, J.  2003, astro-ph/0304495

\bibitem{cm03}
	Cordes, J.~M.\ \& McLaughlin, M.~A.  2003, ApJ, 596, 1142

\bibitem{cl02}
	Cordes, J.~M.\ \& Lazio, T.~J.~W.  2002, astro-ph/0207156

\bibitem{cls97}
	Cordes, J.~M., Lazio, T.~J.~W., \& Sagan, C.\ 1997, ApJ, 487,
	782

\bibitem{cl91}	
	Cordes, J.~M.\ \& Lazio, T.~J.  1991, ApJ, 376, 123

\bibitem{cmmcis81}
	Cortiglioni, S., Mandolesi, N., Morigi, G., Ciapi, A., Inzani,
	P., \& Sironi, G.  1981, Astrophys. \& Space Sci., 75, 153

\bibitem{dkhw99}
	Dav{\'e}, R., Hernquist, L., Katz, N., \& Weinberg, D.~H.
	1999, ApJ, 511, 521

\bibitem{daveetal01}
	Dav{\'e}, R., et al.  2001, ApJ, 552, 473

\bibitem{dl70}
	Davies, J.~G., \& Large, M.~I.  1970, MNRAS, 149, 301

\bibitem{dr85}
	Desch, M.~D.\ \& Rucker, H.~O.	1985, 
	Adv.\ Space Res., 5, 333

\bibitem{ehm74}
	Edwards, P.~J., Hurst, R.~B., \& McQueen, M.~P.~C.  1974,
	Nature, 247, 444

\bibitem{esamdinetal04}
	Esamdin, A., et al.  2004, MNRAS, submitted

\bibitem{fg03}
	Falcke, H.\ \& Gorham, P.  2003, Astroparticle Phys., 19, 477

\bibitem{fdz99}
	Farrell, W.~M., Desch, M.~D., \& Zarka, P.  1999, JGR, 104, 14025

\bibitem{fbbwpgf97}
	Fender, R.~P., Bell Burnell, S.~J., Waltman, E.~B., Pooley,
	G.~G., Ghigo, F.~D., \& Foster, R.~S.  1997, MNRAS, 288, 849

\bibitem{gd81}
	Gallagher, D.~L.\ \& D'angelo, N.  1981, Geophys.\ Res.\ Lett.,
	8, 1087

\bibitem{g-spr03}
	Garc{\'\i}a-S{\'a}nchez, J., Paredes, J.~M., \& Rib\'o, M.
	2003, A\&A, 403, 613

\bibitem{gs66}
	Goldreich, P.\ \& Soter, S.  1966, Icarus, 5, 375 

\bibitem{g97} Goodman, J.  1997, New Astron., 2, 449

\bibitem{g74}
	Gurnett, D.~A.	1974, 
	J.\ Geophys.\ Res., 79, 4227

\bibitem{hkwe03}
	Hankins, T.~H., et al. 2003, Nature, 422, 141

\bibitem{heo96}
	Hankins, T.~H., Ekers, R.~D., \& O'Sullivan, J.~D.
	1996, MNRAS, 283, 1027

\bibitem{hankinsetal81}
	Hankins, T.~H., et al.  1981, ApJ, 244, L61

\bibitem{hr75}
	Hankins, T.~H.\ \& Rickett, B.~J.\ 1975, Methods in
	Computational Physics.\ Volume 14 - Radio astronomy, 14, 55

\bibitem{h71} Hankins, T.~H.\ 1971, ApJ, 169, 487

\bibitem{hl01}
	Hansen, B.~M.~S.\ \& Lyutikov, M.  2001, MNRAS, 322, 695

\bibitem{hw74}
	Hesse, K.~H.\ \& Wielebinski, R.  1974, A\&A, 31, 409

\bibitem{hb-bpsc68}
	Hewish, A., Bell-Burnell, J., Pilkington, J.~D.~H., Scott,
	P.~F., \& Collins, R.~A.  1968, Nature, 217, 709

\bibitem{hf03}
	Huege, T.\ \& Falcke, H.  2003, A\&A, 412, 19

\bibitem{hr73}
	Hughes, V.~A.\ \& Retallack, D.~S.  1973, Nature, 242, 105

\bibitem{hm74}
	Huguenin, G.~R.\ \& Moore, E.~L.  1974, ApJ, 187, L57

\bibitem{ismm82}
	Inzani, P., Sironi, G., Mandolesi, N., \& Morigi, G.  1982,
	AIP Conf.\ Proc.\ 77: Gamma Ray Transients and Related
	Astrophysical Phenomena, eds.\ R.~E.~Lingenfelter,
	H.~S.~Hudson, D.~M.~Worrall (New York: AIP) p.~79

\bibitem{jkw89}
	Jackson, P.~D., Kundu, M.~R., \& White, S.~M.  1989, A\&A,
	210, 284

\bibitem{jr03}
	Johnston, S.\ \& Romani, R.~W. 2003, ApJ, 590, L95

\bibitem{kkhm04}
	Karachentsev, I.~D., Karachentseva, V.~E., Huchtmeier, W.~K.,
	\& Makarov, D.~I.  2004, AJ, 127, 2031

\bibitem{kardashevetal77}
	Kardashev, N.~S., et al.  1977, Soviet Astron., 21, 1

\bibitem{khcm03}
	Katz, C.~A., Hewitt, J.~N., Corey, B.~E., \& Moore, C.~B.
	2003, PASP, 115, 675

\bibitem{k-cjwtr01}
	Kedziora-Chudczer, L.~L., Jauncey, D.~L., Wieringa, M.~H.,
	Tzioumis, A.~K., \& Reynolds, J.~E.  2001, MNRAS, 325, 1411

\bibitem{kgwwp94}
	Koranyi, D.~M., Green, D.~A., Warner, P.~J., Waldram, E.~M.,
	\& Palmer, D.~M.  1994, MNRAS, 271, 51

\bibitem{kgwwp95}
	Koranyi, D.~M., Green, D.~A., Warner, P.~J., Waldram, E.~M.,
	\& Palmer, D.~M.  1995, MNRAS, 276, L13

\bibitem{krameretal04}
	Kramer, M., et al.  2004, in prep.

\bibitem{l94}
	Lainela, M.\ 1994, A\&A, 286, 408

\bibitem{lmdekz00}
	Langston, G., Minter, A., D'Addario, L., Eberhardt, K., Koski,
	K., \& Zuber, J.  2000, AJ, 119, 2801

\bibitem{lazioetal04}
	Lazio, T.~J.~W., Farrell, W.~M., et al.  2004, ApJ, in press

\bibitem{lc03}
	Lazio, T.~J.~W.\ \& Cordes, J.~M.  2003, ``Telescope Response
	Times,'' SKA Memorandum~31

\bibitem{lbrbmk98}
	Lecacheux, A., Boudjada, M.~Y., Rucker, H.~O., Bougeret,
	J.~L., Manning, R., \& Kaiser, M.~L.\ 1998, A\&A, 329, 776

\bibitem{le80}
	Linscott, I.~R.\ \& Erkes, J.~W.  1980, ApJ, 236, L109

\bibitem{ljbk-cmrt03}
	Lovell, J.~E.~J., Jauncey, D.~L., Bignall, H.~E.,
	Kedziora-Chudczer, L., Macquart, J.-P., Rickett, B.~J., \&
	Tzioumis, A.~K.  2003, AJ, 126, 1699

\bibitem{lcumlfh95} 
	Lundgren, S.~C., Cordes, J.~M., Ulmer, M., Matz, S.~M.,
	Lomatch, S., Foster, R.~S., \& Hankins, T.  1995, ApJ, 453,
	433

\bibitem{mkcasms96}
	Mann, G., Klassen, A., Classen, H.-T., Aurass, H., Scholz, D.,
	MacDowall, R.~J., \& Stone, R.~G.  1996, A\&AS, 119, 489

\bibitem{m04}
	Marcy, G.~W.  2004,
	in Bioastronomy 2002: Life Among the Stars, eds.\ R.~Norris,
	C.~Oliver, \& F.~Stootman (Astron.\ Soc.\ Pacific: San
	Francisco) in press

\bibitem{mclaughlinetal04}
	McLaughlin, M.~A., et al.  2004, in preparation

\bibitem{mc03}
	McLaughlin, M.~A.\ \& Cordes, J.~M.  2003, ApJ, 596, 982

\bibitem{megr81}
	McCulloch, P.~M., Ellis, G.~R.~A., Gowland, G.~A., \& Roberts,
	J.~A.  1981, ApJ, 245, L99

\bibitem{mitchelletal01}
	Mitchell, D.~L., Lin, R.~P., Mazelle, C., et al.  2001,
	J.\ Geophys.\ Res.-Planets, 106, 23419

\bibitem{n03}
	Nemiroff, R.~J.  2003, AJ, 125, 2740

\bibitem{nft95}
	Nice, D.~J., Fruchter, A.~S., \& Taylor, J.~H.  1995, ApJ,
	449, 156

\bibitem{n99}
	Nice, D.~J.  1999, ApJ, 513, 927

\bibitem{oes78}
	O'Sullivan, J.~D., Ekers, R.~D., \& Shaver, P.~A.  1978,
	Nature, 276, 590

\bibitem{pt79} Phinney, S.\ \& Taylor, J.~H.  1979, Nature, 277, 117

\bibitem{pscdm88}
	Poquerusse, M., Steinberg, J.~L., Caroubalos, C., Dulk, G.~A.,
	\& MacQueen, R.~M.  1988, A\&A, 192, 323

\bibitem{r77}
	Rees, M.~J.\ 1977, Nature, 266,333

\bibitem{rwgr03}
	Richards, M.~T., Waltman, E.~B., Ghigo, F.~D., \& Richards,
	D.~St.~P.  2003, ApJS, 147, 337

\bibitem{rj01} Romani, R.~W.~\& Johnston, S.  2001, ApJ, 557, L93

\bibitem{sw02}
	Sagiv, A.\ \& Waxman, E.   2002, ApJ, 574, 861

\bibitem{s03}
	Schneider, J.  2003, ``Extrasolar Planets Encyclopaedia,''
	\hfil\\ 
	http://www.obspm.fr/encycl/encycl.html

\bibitem{swb03}
	Shkolnik, E., Walker, G.~A.~H., \& Bohlender, D.~A.
	2003, ApJ, 597, 1092

\bibitem{sr68}
	Staelin, D.~H.\ \& Reifenstein, E.~C.  1968, Science, 162,
	1481

\bibitem{tbd81}
	Taylor, J.~H., Backus, P.~R., \& Damashek, M.  1981, ApJ, 244, L65

\bibitem{thh72}
	Taylor, J.~H., Huguenin, G.~R., \& Hirsch, R.~M.  1972, ApJ,
	172, L17

\bibitem{t00}
	Trilling, D.~E. 2000, ApJ, 537, L61

\bibitem{tc-wbjk87}
	Turtle, A.~J., Campbell-Wilson, D., Bunton, J.~D., Jauncey,
	D.~L., \& Kesteven, M.~J.  1987, Nature, 327, 38

\bibitem{uk00}
	Usov, V.~V.\ \& Katz, J.~I.  2000, A\&A, 364, 655

\bibitem{vl89}
	Vaughan, A.~E.\ \& Large, M.~I.  1989, ApJ, 25, 159

\bibitem{wgjffs95}
	Waltman, E.~B., Ghigo, F.~D., Johnston, K.~J., Foster, R.~S.,
	Fiedler, R.~L., \& Spencer, J.~H.  1995, AJ, 110, 290

\bibitem{w69} Weber, J. 1969, Phys.\ Rev.\ Letters, 22, 1320

\bibitem{w01}
	Weekes, T.  2001, in Radio Detection of High Energy Particles,
	First International Workshop RADHEP 2000, eds.\ D.~Saltzberg
	\& P.~Gorham (Melville, NY: AIP) p.~3

\bibitem{wksysk03}
	White, S.~M., Krucker, S., Shibasaki, K., Yokoyama, T.,
	Shimojo, M., \& Kundu, M.~R. 2003, ApJ, 595, L111

\bibitem{wl79}
	Wu, C.~S.\ \& Lee, L.~C.  1979, 
	ApJ, 230, 621

\bibitem{y86}
	Yudaeva, N.~A.  1986, Soviet Astron.\ Letters, 12, 150

\bibitem{ztrr01}
	Zarka, P., Treumann, R.~A., Ryabov, B.~P., \& Ryabov, V.~B.
	2001, Ap\&SS, 277, 293

\bibitem{z04} 
	Zarka, P., et al.  2004, Planetary Space Sci., in press

\end{thebibliography}
\end{document}